\documentclass[preprint,aps,prd,showpacs,nofootinbib]{revtex4-1}

\usepackage{amsmath}
\usepackage{amsfonts}
\usepackage{mathrsfs}
\usepackage{color}

\newcommand{\cL}{\mathcal{L}}
\newcommand{\cM}{\mathcal{M}}

\newcommand{\cO}{\mathcal{O}}

\newcommand{\cR}{\mathcal{R}}

\newcommand{\cU}{\mathcal{U}}
\newcommand{\cV}{\mathcal{V}}

\newcommand*{\ie}{\textit{i.e.},\ }

\newcommand*{\et}{\textit{et al.}}
\newcommand*{\pc}{\textit{p.c.}}

\newcommand*{\npb}[1]{Nucl.\ Phys.\ \textbf{B#1}}
\renewcommand*{\prd}[1]{Phys.\ Rev.\ \textbf{D#1}}

\newcommand*{\npbps}[1]{Nucl.\ Phys.\ \textbf{B}
    (Proc.\ Suppl.) \textbf{#1}}

\newcommand*{\chpt}{\raise0.4ex\hbox{$\chi$}PT}
\newcommand*{\schpt}{S\raise0.4ex\hbox{$\chi$}PT}
\newcommand*{\rschpt}{rS\raise0.4ex\hbox{$\chi$}PT}
\newcommand*{\pqschpt}{PQ-S\raise0.4ex\hbox{$\chi$}PT}
\newcommand*{\pqrschpt}{PQ-rS\raise0.4ex\hbox{$\chi$}PT}
\newcommand*{\Tr}{\textrm{Tr}}
\newcommand*{\str}{\textrm{Tr}}
\newcommand{\pamu}{\partial_\mu}

\newcommand{\SigmaDag}{\Sigma^{\dagger}}
\newcommand{\chiDag}{\chi^{\dagger}}
\newcommand{\pmuSigma}{\pamu \Sigma}

\newcommand{\pmuSigmaDag}{\pamu \SigmaDag}

\newcommand{\dmuSigma}{D_\mu\Sigma}
\newcommand{\dnuSigma}{D_\nu \Sigma}
\newcommand{\dmuSigmaDag}{D_\mu \SigmaDag}
\newcommand{\dnuSigmaDag}{D_\nu \SigmaDag}

\newcommand{\nr}{n_r}
\newcommand{\nrp}{n'_{r}}
\newcommand{\hermi}{\it{h.c.}}
\newcommand{\ximu}{\xi_\mu}
\newcommand{\xifivetwo}{\xi_5^{(R)}}
\newcommand{\xinutwo}{\xi_{\nu}^{(R)}}
\newcommand{\xinufivetwo}{\xi_{\nu 5}^{(R)}}
\newcommand{\xifivenutwo}{\xi_{5\nu}^{(R)}}
\newcommand{\ximunutwo}{\xi_{\mu\nu}^{(R)}}
\newcommand{\xinumutwo}{\xi_{\nu \mu}^{(R)}}

\newcommand{\fpi}{f_\pi}
\newcommand{\mutwo}{\mu_{(2)}}
\newcommand{\ftwo}{f_{(2)}}
\newcommand{\deltapvtwo}{{\delta'_V}^{(2)}}
\newcommand{\deltapatwo}{{\delta'_A}^{(2)}}
\newcommand{\Lptwo}{{L'}_{(2)}}
\newcommand{\Lpptwo}{{L''}_{(2)}}
\newcommand{\TildeLpptwo}{\tilde L''_{(2)}}
\newcommand{\TildeLptwo}{\tilde L'_{(2)}}
\newcommand{\TildeLpptwob}{\tilde L^{''0}_{(2)}}
\newcommand{\TildeLptwob}{\tilde L^{'0}_{(2)}}

\newcommand{\denom}{16 \pi^2 \ftwo^2}
\newcommand{\Lambdafactor}{\Lambda^{d-4}}
\newcommand{\mjpower}{m_j^2 (m_j^2)^{-\frac{\epsilon}{2}}}

\newcommand{\deltapv}{\delta'_V}
\newcommand{\deltapa}{\delta'_A}
\newcommand{\mxvsq}{m_{X_V}^2}

\newcommand{\muvsq}{m_{U_V}^2}
\newcommand{\muasq}{m_{U_A}^2}

\newcommand{\mxisq}{m_{X_I}^2}
\newcommand{\myisq}{m_{Y_I}^2}
\newcommand{\muisq}{m_{U_I}^2}

\newcommand{\Cc}{\frac{1}{16\pi^2}}

\newcommand{\logmk}{\log\frac{\mu m_s}{\Lambda^2}}
\newcommand{\logmeta}{\log\frac{\frac{4}{3}\mu m_s}{\Lambda^2}}

\begin{document}

\title{Staggered chiral perturbation theory in the two-flavor case}
\author{Xining Du}
\affiliation{Department of Physics, Washington University, St.Louis, Missouri,
USA}
%\email{xiningdu@physics.wustl.edu}
\begin{abstract}
I study two-flavor staggered chiral perturbation theory in the
light pseudoscalar sector. The pion mass and decay constant are calculated
through NLO in the partially-quenched case. In the limit where the strange quark
mass is large compared to the light quark masses and the taste splittings, I
show that the SU(2) staggered chiral theory emerges from the SU(3) staggered
chiral theory, as expected. Explicit relations between SU(2) and SU(3)
low energy constants and taste-violating parameters are given. The results are
useful for SU(2) chiral fits to asqtad data and allow one to incorporate effects
from
varying strange quark masses.
\end{abstract}

\maketitle

\section{INTRODUCTION}

Chiral perturbation theory\ (\chpt)~\cite{WEINBERG, GL_SU3, GL_SU2}  has proved
to be a very important tool for
lattice QCD simulations. By using \chpt, one can extrapolate physical quantities
to physical light quark masses as well as getting information on low energy
constants\ (LECs) in the chiral theory. Nowadays many lattice simulations use
2+1 flavor dynamical quarks, and SU(3) \chpt\  is often used for the chiral
analysis.
However, SU(2) \chpt\ is of interest for the following reasons:
\begin{itemize}
\item The strange quark mass in lattice simulations is usually very close
to its physical value, while light quark masses are significantly
smaller. Hence we expect that SU(2) \chpt\ converges faster than SU(3) \chpt\
and
serves as a better approximation.
\item By fitting lattice data to SU(2) \chpt\
formulae, one can extract information about SU(2) LECs directly and compare them
with results from phenomenological analysis.
\item By comparing results from SU(2) and SU(3) chiral fits, we can study the
systematic errors resulting from the truncations of each version of \chpt.
\end{itemize}
The use of SU(2) \chpt\ for chiral fits to data from three-flavor simulations 
has been advocated by several groups recently~\cite{HEAVYs_RBC, HEAVYs_PACS-CS}; 
see also the review by Lellouch~\cite{LELLOUCH} and references therein.

For lattice QCD formulated with (rooted) staggered fermions, rooted staggered
chiral perturbation
theory\ (\rschpt)~\cite{LEE_SHARPE, AB_M, CB_SCHPT, BGS_STAGGERED} is the
corresponding effective
field theory (EFT), which incorporates taste-violating effects systematically.
The partially-quenched version,
\pqrschpt, was used by the MILC collaboration in the SU(3) chiral analysis of
lattice
data generated with 2+1 asqtad fermions. This was done at NLO systematically by
using 1-loop chiral logarithms from SU(3) \pqrschpt\ plus higher order analytic
terms~\cite{MILC_04}. Recently, a systematic NNLO analysis~\cite{URS_LAT09} was
performed with continuum NNLO chiral logarithms. Results from
this work give strong evidence for the validity of of \rschpt\ as the EFT for
QCD formulated with rooted staggered quarks.

In order to perform the corresponding analysis in the SU(2) case, one needs to
calculate the 1-loop formulae for pseudoscalar meson masses and decay constants
in two-flavor \pqrschpt. In addition, it is important to check that the presence
of taste violations and rooting do not interfere with the decoupling of the
strange quark as its mass is increased, allowing the SU(2) chiral theory to
emerge from the SU(3) theory. This is a check on a technical step in the
argument of Ref.~\cite{CB_SCHPT} that \rschpt\ is the correct effective chiral
theory for rooted staggered quarks. Finally, it is useful to relate the LECs in
the
two-flavor and three-flavor cases, and to find the scale dependence of the LECs
in both cases, thereby checking their consistency.
These calculations are presented below.

\section{TWO-FLAVOR PARTIALLY-QUENCHED ${\text{\rschpt}}$ }

\subsection{Brief review of \schpt}

The key point of \schpt\ is to incorporate systematically the taste-violating 
effects at finite lattice spacing in the chiral perturbation theory for
staggered fermions. The idea of how to develop \chpt\ including scaling
violations is due to Sharpe and Singleton~\cite{SHARPE_SINGLETON}, and was 
first applied to staggered quarks by Lee and Sharpe~\cite{LEE_SHARPE}.

Basically, \schpt\ is constructed through two steps. First, one writes down the
continuum Symanzik Effection Theory (SET) for staggered fermions. The taste-violating 
four-quark operators appear at $\cO(a^2)$ in the SET. The coefficient of each of these 
operators also depends on the coupling constant $\alpha_s$, and
it varies with different staggered actions used in simulations. Specifically,
for unimproved staggered action, these operators appear at $\cO(\alpha_s a^2)$, while for 
asqtad improved action, these operators appear at $\cO(\alpha_s^2
a^2)$~\cite{MILC_04}. For Highly Improved Staggered Quarks (HISQ), these
operators also appear at $\cO(\alpha_s^2 a^2)$ but with smaller coefficients
than for asqtad quarks~\cite{HISQ, MILC_HISQ08}. In the
second step, one maps operators in the SET to terms in the chiral Lagrangian
using spurion analysis. The taste-violating four quark operators are mapped into 
the taste-breaking potential in the chiral Lagrangian. In the two-flavor case, 
these two steps can be done in the same manner as those in the three-flavor case 
given by Refs.~\cite{AB_M}. The final form of the two-flavor chiral Lagrangian 
looks exactly the same as the three-flavor Lagrangian except that the chiral field 
$\Phi$ takes its definition in the two-flavor case.

For the purposes of constructing the chiral theory, the SET is taken as
``given". We do not need to consider the issues of additive and multiplicative
renormalizations that one would need to face in defining finite higher
dimensional operators in perturbation theory. All we need to know are the
symmetry properties of staggered fermions, which determine what operators can
appear. Note further that the lattice spacing $a$ is not a cutoff for the chiral
theory, which will in practice be cut off using dimensional regularization.
Instead $a$ serves to parameterize symmetry breaking in the chiral theory,
and plays a role closely analogous to that of the light quark masses.

In the SET there are also operators at $\cO(a^2)$ which satisfy all the 
continuum symmetries of staggered fermions. Such operators produce ``generic"
discretization effects and in general come with different powers of $\alpha_s$
than taste-violations. (For example, with asqtad quarks, the lowest order of generic 
discretization corrections is $\cO(\alpha_s a^2)$ while the lowest order of taste-violations 
is $\cO(\alpha_s^2 a^2)$.) These operators in the SET are logically distinct from $\cO(a^2)$ 
taste-violating operators, and their sizes are ``dialed" more or less independently by 
adjustments of the actions. In the asqtad case, it is known from simulations~\cite{MILC_04} that taste
violating effects are the dominant cause of discretization effects at
$\cO(a^2)$ even though generic effects can appear at lower order in $\alpha_s$. That
is because the coefficients of the taste-violating operators turn out to be large.
After being mapped to chiral theory, the generic SET operators give the same terms as those in the 
continuum Lagrangian, but multiplied by a coefficient of $\cO(a^2)$. 
For the same reason above, these terms in the chiral Lagrangian representing generic
discretization effects are essentially different from the taste-violating terms
even though both of them can appear at the same order of lattice spacing $a$. 
It is therefore consistent to consider the effects of taste-violating
operators independently of generic effects, and that is what I do here. 

In practical numerical work, both effects need to be considered. The fact that
taste-violations and generic finite lattice spacing effect usually are significantly 
small, have different mass dependence, and come with different powers of $\alpha_s$ allows a
relatively clean separations if sufficient numbers of different lattice spacings
are included. Of course, some systematic error will be present and needs to be
estimated.

For convenience in numerical work, the effects of generic operators are often
absorbed into effective $a^2$ dependence of the LECs. This is possible since the
generic operators have the same symmetries as the continuum QCD
operators.\footnote{There are also operators that have continuum taste symmetry
but violate rotational invariance. Their effects appear only at $\cO(a^4)$ in
the \chpt\ for pseudoscalar mesons.} So, for example, one can take the results
given here for SU(2) \schpt\ and effectively take into account generic operators
simply by letting the LECs have $a^2$ dependence. But I emphasize that,
logically, the generic effects should be thought of in \chpt\ as new operators,
just like the taste-violating effects, not as corrections to old operators. That 
way, we satisfy the requirement that all LECs in \schpt\ are $a^2$ independent,
just as they are independent of the light quark masses.

\subsection{Two-flavor PQ-\schpt\ at LO}

In \schpt\, the theory becomes a joint expansion about the chiral and continuum limits. 
The effective Lagrangian was worked out for the single flavor case in Ref.~\cite{LEE_SHARPE},
and later generalized to multi-flavor case in Ref.~\cite{AB_M}.
In Refs.~\cite{CB_SCHPT, BGS_STAGGERED}, it was shown that the replica method
introduced for this problem in Ref.~\cite{AB_NPB} is a valid method for taking
rooting into account. The partial quenching can be treated either by the graded
symmetry method~\cite{BG_92, BG_93}, or by the replica method~\cite{DS_REPLICA}.
Here, for simplicity, I use the replica method for both the rooting and the
partial quenching.
I take $\nrp$ copies of each valence quark (x,y), and $\nr$ copies of each
flavor of sea quark (u,d).
The chiral symmetry group is $SU(8(\nrp + \nr))_L \times SU(8(\nrp + \nr))_R$.
The pseudoscalar mesons can now be collected into a $8(\nrp+\nr) \times 8 (\nrp
+ \nr)$ matrix $\Phi$, where the factors of 8 arise from 2 flavors of 4 tastes
each:
\begin{equation}
        \Phi = \left( \begin{array}{cccccccccccc}
X^{11}      & \ldots    & X^{1\nrp}     & P_+^{11}     & \ldots & P_+^{1
\nrp}   & \ldots & \ldots & \ldots & \ldots & \ldots & \ldots \\*
\vdots      & \ddots    & \vdots        & \vdots       & \ddots & \vdots       &
\ldots & \ldots & \ldots & \ldots & \ldots & \ldots \\*
X^{\nrp 1}  & \ldots    & X^{\nrp\nrp}  & P_+^{\nrp 1} & \ldots &
P_+^{\nrp\nrp} & \ldots & \ldots & \ldots & \ldots & \ldots & \ldots \\*
P_-^{11}     & \ldots    & P_-^{1\nrp}   & Y^{11}     & \ldots & Y^{1 \nrp}
&\ldots & \ldots & \ldots & \ldots & \ldots & \ldots \\*
\vdots      & \ddots    & \vdots        & \vdots     & \ddots & \vdots       &
\ldots & \ldots & \ldots & \ldots & \ldots & \ldots \\*
P_-^{\nrp 1}& \ldots    & P_-^{\nrp\nrp}& Y^{\nrp 1} & \ldots & Y^{\nrp\nrp}
&  \ldots & \ldots & \ldots & \ldots & \ldots & \ldots \\*
\vdots      & \vdots    & \vdots        & \vdots     & \vdots & \vdots
&U^{11}   & \ldots & U^{1\nr}   & \pi_+^{11}     & \ldots & \pi_+^{1 \nr}
\\*
\vdots      & \vdots    & \vdots        & \vdots     & \vdots & \vdots
&\vdots   & \ddots & \vdots     & \vdots         & \ddots & \vdots \\*
\vdots      & \vdots    & \vdots        & \vdots     & \vdots & \vdots
&U^{\nr 1}   & \ldots & U^{\nr\nr}   & \pi_+^{\nr1}     & \ldots &
\pi_+^{\nr\nr} \\*
\vdots      & \vdots    & \vdots        & \vdots     & \vdots & \vdots
&\pi_-^{11}   & \ldots & \pi_-^{1\nr}   & D^{11}     & \ldots & D^{1 \nr}
\\*
\vdots      & \vdots    & \vdots        & \vdots     & \vdots & \vdots
&\vdots   & \ddots & \vdots     & \vdots         & \ddots & \vdots \\*
\vdots      & \vdots    & \vdots        & \vdots     & \vdots & \vdots
&\pi_-^{\nr 1}   & \ldots & \pi_-^{\nr\nr}   & D^{\nr1}     & \ldots &
D^{\nr\nr}
                      \end{array} \right) ,
\end{equation}
where each entry is a $4\times 4$ matrix in taste space with, for example,
$U^{ij}=\sum_{a=1}^{16} U_a^{ij} T_a$. $X, Y, U$, and $D$ are the mesons made
from $x\bar x, y\bar y, u\bar u$, and $d\bar d$ quarks respectively. $P_+$ is a
charged valence meson made from $x\bar y$\; and $\pi_+$ is the charged sea meson
made from $u\bar d$. The hermitian generators $T_a$ are defined to be:
\begin{align}
T_a = \{\xi_5, i\xi_{\mu 5}, i\xi_{\mu\nu}, \xi_{\mu}, \xi_I\}.
\end{align}
The lowest order ($\cO(p^2,m_q,a^2)$) Euclidean Lagrangian is:
\begin{align}
\cL^{(2)} = &\frac{\ftwo^2}{8} \Tr(\dmuSigma\dmuSigmaDag)
   - \frac{\ftwo^2}{8}\Tr(\chi\SigmaDag + \chi\Sigma) \nonumber \\*
      &+\frac{2m_0^2}{3}(U_I^{11} + \ldots + U_I^{\nr\nr} +D_I^{11} + \ldots +
D_I^{\nr\nr})^2 + a^2\cV, \label{eq:LagLO}
\end{align}
where $\Sigma=\exp(i\Phi/f)$ and $\chi$ is a $8(\nrp + \nr) \times 8(\nrp + \nr)$ diagonal matrix:
\begin{equation}
\chi = 2\mutwo Diag( \underbrace{m_x I, \ldots, m_x I}_{\nrp}, \underbrace{m_y
I, \ldots, m_y I}_{\nrp}, \underbrace{m_u I, \ldots, m_u I}_{\nr},
\underbrace{m_d I, \ldots, m_d I}_{\nr})
\end{equation}
with $I$ the $4\times 4$ identity matrix in taste space. The covariant
derivative $D_\mu$ in Eq.~(\ref{eq:LagLO}) is defined by
\begin{equation}
\dmuSigma = \pmuSigma - il_\mu\Sigma + i\Sigma r_\mu, \qquad \dmuSigmaDag =
\pmuSigmaDag - ir_\mu\SigmaDag + i\SigmaDag l_\mu,
\end{equation}
where $l_\mu$ and $r_\mu$ are the left and right-handed currents respectively.
Throughout this paper, I alway use the superscript or subscript ``(2)" to indicate
parameters in the two-flavor theory.

The taste-breaking potential $\cV = \cU + \cU'$ is defined by:
\begin{align}
    -\cU \equiv \sum_k C_k \cO_k = &C_1^{(2)} \Tr(\xifivetwo \Sigma \xifivetwo
\SigmaDag) \nonumber \\
                &+ C_3^{(2)} \frac{1}{2}\sum_\nu [\Tr(\xinutwo \Sigma
\xinutwo \Sigma) + \hermi] \nonumber \\
                &+ C_4^{(2)} \frac{1}{4}\sum_\nu [\Tr(\xinufivetwo
\Sigma \xifivenutwo \Sigma)  + \hermi] \nonumber \\
                &+ C_6^{(2)} \sum_{\mu < \nu} \Tr(\ximunutwo \Sigma
\xinumutwo \SigmaDag), \\
    -\cU' \equiv \sum_{k'}C_{k'}\cO_{k'} = &C_{2V}^{(2)}\frac{1}{4}\sum_\nu
[\Tr(\xinutwo\Sigma) \Tr(\xinutwo\Sigma) + \hermi] \nonumber \\
                        &+C_{2A}^{(2)}\frac{1}{4}\sum_\nu
[\Tr(\xinufivetwo \Sigma) \Tr(\xifivenutwo \Sigma) + \hermi] \nonumber \\
                        &+C_{5V}^{(2)}\frac{1}{2}\sum_\nu
[\Tr(\xinutwo \Sigma) \Tr(\xinutwo \SigmaDag)] \nonumber \\
                        &+C_{5A}^{(2)}\frac{1}{2}\sum_\nu[\Tr(\xinufivetwo
\Sigma) \Tr(\xifivenutwo\SigmaDag)],
\end{align}
where $\xifivetwo$ is the product of $\xi_5$ in taste space with the identity
matrix in flavor and replica space, and similarly for $\xinutwo, \xinufivetwo$ and
$\ximunutwo$. 

Due to the anomaly, the SU($8(\nrp+\nr)$) singlet receives a large
contribution to its
mass ($\propto m_0$), and thus does not play a dynamical role. Integrating out
this
singlet is equivalent to keeping the singlet explicitly in the
Lagrangian (the third term in $\cL^{(2)}$), and taking $m_0 \to \infty$ at the
end of
the calculation~\cite{SHARPE_SHORESH}. Here, the $m_0^2$ term is normalized so
that for the hairpin diagram between two flavor-neutral taste singlet
mesons, each composed of a single species, the vertex is $\frac{4m_0^2}{3}$,
independent of the number of flavors.
For the two-flavor \schpt\ with $\nr$ replicas for each sea quark, the mass
matrix for flavor-neutral taste singlet mesons takes the form:
\begin{equation}
 \left( \begin{array}{cccccc}
m_{U_I^{11}}+\delta' & \delta' & \delta' & \delta' & \ldots & \delta' \\
\delta'              & \ddots  & \delta' & \vdots  & \ddots & \vdots \\
\delta'              & \ldots  & m_{U_I^{\nr\nr}}+\delta' &  \delta' & \ldots &
\delta'\\
\delta' & \ldots & \delta' & m_{D_I^{11}}+\delta' & \delta' & \delta' \\
\vdots  & \ddots & \vdots  & \delta'      & \ddots  & \delta' \\
\delta' & \ldots & \delta' & \delta'      & \delta' & m_{D_I^{\nr\nr}}+\delta'    
        \end{array} \right),    
\end{equation}
where every non-diagonal element is $\delta' \equiv \frac{4m_0^2}{3}$, and I
have anticipated taking $\nrp \to 0$ to eliminate virtual loops of valence
quarks. Diagonalizing the matrix and taking the limit of $m_0 \to \infty$, we
obtain the
mass of the $\eta'_I$:
\begin{equation}
m^2_{\eta'_I} = \frac{8m_0^2}{3} \nr.
\end{equation}
Generally, if there are $N_f$ flavors of sea quarks, the result will be
$\frac{4m_0^2}{3} N_f \nr$.

\subsection{Two-flavor PQ-\rschpt\ at NLO}

At NLO, the two-flavor PQ-\rschpt\ Lagrangian has two parts:
\begin{equation}
\cL^{(4)} = \cL_{cont}^{(4)} + \cL_{\operatorname{t-v}}^{(4)}. \label{eq:LagNLO}
\end{equation}
$\cL_{cont}^{(4)}$ contains operators of $\cO(p^4, p^2 m_q, m_q^2)$, which are
of the same form as operators in two-flavor continuum PQ-\chpt.
$\cL_{\operatorname{t-v}}^{(4)}$
is of $\cO(a^2 p^2, a^2 m_q, a^4)$. It contains all NLO taste-violating terms
for staggered fermions~\cite{SHARPE_WATER}.

The most general continuum NLO Lagrangian $\cL_{cont}^{(4)}$ in Euclidean space 
can be written as:
\begin{align}
    \cL_{cont}^{(4)} &= -\frac{l_1^0}{4}[\Tr (\dmuSigmaDag \dmuSigma)]^2 -
\frac{l_2^0}{4}
    \Tr(\dmuSigmaDag \dnuSigma) \Tr(\dmuSigmaDag \dnuSigma) \nonumber \\*
    &+ p_3^0 \Big( \Tr(\dmuSigmaDag\dmuSigma \dnuSigmaDag\dnuSigma) -
\frac{1}{2}
    [\Tr(\dmuSigmaDag\dmuSigma)]^2 \Big)\nonumber \\*
        &+ p_4^0 \Big( \Tr(\dmuSigmaDag\dnuSigma\dmuSigmaDag\dnuSigma) +
2\Tr(\dmuSigmaDag\dmuSigma\dnuSigmaDag\dnuSigma) \nonumber \\*
        &- \frac{1}{2}[\Tr(\dmuSigmaDag\dmuSigma)]^2 -
\Tr(\dmuSigmaDag\dnuSigma)
\Tr(\dmuSigmaDag\dnuSigma) \Big) \nonumber \\*
        &- \frac{l_3^0 + l_4^0}{16}[\Tr(\chi\SigmaDag + \Sigma\chiDag)]^2 +
\frac{l_4^0}{8}\Tr(\dmuSigmaDag\dmuSigma)\Tr(\chi\SigmaDag + \Sigma\chiDag)
\nonumber \\*
        &+ \frac{p_1^0}{16} \Big(\Tr(\dmuSigmaDag\dmuSigma(\chi\SigmaDag +
\Sigma\chiDag)) -
\frac{1}{2}\Tr(\dmuSigmaDag\dmuSigma)\Tr(\chi\SigmaDag + \Sigma\chiDag)
\Big)\nonumber \\*
       &+ \frac{p_2^0}{16} \Big( 2 \Tr(\SigmaDag\chi\SigmaDag\chi +
\Sigma\chiDag\Sigma\chiDag) -
\Tr(\chi\SigmaDag+ \Sigma\chiDag)^2 - \Tr(\chi\SigmaDag - \Sigma\chiDag)^2
\Big) \nonumber \\
       &+ \frac{l_7^0}{16} [\Tr(\chi\SigmaDag - \Sigma\chiDag)]^2\nonumber \\*
       & - l_5^0 \Tr(\SigmaDag{F_R}_{\mu\nu}\Sigma{F_L}_{\mu\nu}) - \frac{i
l_6^0}{2}\Tr({F_L}_{\mu\nu}\dmuSigmaDag\dnuSigma +
{F_R}_{\mu\nu}\dmuSigma\dnuSigmaDag) \nonumber \\
       &+ \mbox{contact terms},
       \label{eq:NLOL}
\end{align}
It is written in this form so that the bare coefficients $l_i^0\
(i=1,2,\cdots,7)$ have the same values
as the corresponding $l_i$ with the standard definitions~\cite{GL_SU2} in
the two-flavor full QCD limit. The parameters $p_1^0$, $p_2^0$, $p_3^0$ and
$p_4^0$ are the four extra LECs at NLO in the partially-quenched case. The four
operators associated
with $p_i^0$ are unphysical operators at $\cO(p^4)$, which only appear in the
two-flavor
partially-quenched theory. These unphysical operators vanish in the unquenched
SU(2)
sector of the PQ theory as a result of the Cayley-Hamilton relations for
2-dimensional matrices.
Among these operators, the two with factors $p_1^0$ and $p_2^0$ will contribute
to the
pion masses and decay constants at NLO. The other two with factors $p_3^0$ and
$p_4^0$ only contribute to the same quantities at NNLO, since they contain four
derivatives.
Here, I am only interested in pion masses and decay constants at NLO,
so $p_1^0$ and $p_2^0$ will enter the calculations below, and $p_3^0$ and
$p_4^0$ are irrelevant.

This set of LECs can be related to the LECs used by Bijnens and
L\"ahde~\cite{BIJ_SU2}
through:
\begin{align}
&p_3^0 = -L_3^{(2pq)} + 2 L_0^{(2pq)},        &p_4^0 &= -L_0^{(2pq)}, \nonumber
\\*
&l_1^0 = 4L_1^{(2pq)} + 2L_3^{(2pq)} - 2L_0^{(2pq)},  &l_2^0 &= 4L_2^{(2pq)} +
4L_0^{(2pq)}, \nonumber \\*
&p_1^0 = 16 L_5^{(2pq)},              &p_2^0 &= -8 L_8^{(2pq)}, \nonumber \\*
&l_3^0 = 16L_6^{(2pq)} + 8L_8^{(2pq)} - 8L_4^{(2pq)} - 4L_5^{(2pq)}, &l_4^0 &=
8L_4^{(2pq)} + 4L_5^{(2pq)}, \nonumber \\*
&l_5^0 = L_{10}^{(2pq)},              &l_6^0 &= -2L_9^{(2pq)}, \nonumber \\*
&l_7^0 = -16L_7^{(2pq)} - 8L_8^{(2pq)}.
\end{align}

The general form of $\cL_{\operatorname{t-v}}^{(4)}$ ($\cO(a^2 p^2, a^2 m_q,
a^4)$) is
given in Ref.~\cite{SHARPE_WATER}. Examples of operators in
$\cL_{\operatorname{t-v}}^{(4)}$ that contribute here are:
\begin{equation}
a^2  \str(\pmuSigmaDag\xi_5\pmuSigma\xi_5), \quad a^2 \str(\ximu\SigmaDag\ximu
\chi^\dag) + \pc ,
\label{eq:NLOtvterms}
\end{equation}
(with $\pc$ indicating parity conjugate) where the first operator contributes
both to pseudo-Goldstone masses and decay
constants at NLO, and the second one only contributes to the pseudo-Goldstone
masses at NLO. From this, it is clear that the taste-violating analytic
contributions to decay constants and masses at NLO are independent. We do not
need any further details from Ref.~\cite{SHARPE_WATER} here, since it is not
currently useful to relate the NLO analytic taste-violating contributions to
parameters in the Lagrangian.

\subsection{Rooting and partial quenching}

In the continuum limit, there are four degenerate taste species for each quark
flavor. We obtain physical results in \rschpt\ by taking the fourth root of each
fermion determinant, which is known as the fourth root procedure. Although it
has been shown that this procedure produces, non-perturbatively, violations
of locality at non-zero lattice spacing~\cite{BGS_NONLOCAL}, work over the last
few
years indicates that locality and universality are restored in the continuum
limit of the lattice theory~\cite{SHAMIR_1, SHAMIR_2}, and that \rschpt\ is the
correct chiral effective theory~\cite{CB_SCHPT, BGS_STAGGERED}, thereby
reproducing continuum \chpt\ in the $a\to 0$ limit. For a recent
review of the fourth-root procedure see Ref.~\cite{MILC_REVIEW} and references
therein.

For calculations in \rschpt, the fourth-root is taken by letting $\nr \to
\frac{1}{4}$ at the end of the calculation~\cite{CB_SCHPT, BGS_STAGGERED}.
Similarly, virtual loops associated with the valence quarks are eliminated by
taking $\nrp \to 0$~\cite{DS_REPLICA}.

\section{PION MASS AND DECAY CONSTANT}

Following the procedures in Ref.~\cite{AB_M}, I calculate the light pseudoscalar
mass and decay constant through NLO\ ($\cO(m_q^2, m_q a^2)$). For simplicity, I
always assume the up and down quark masses are equal, $m_u = m_d = m_l$.
The dimensional regularization scheme is employed, and the results in
$d=4-\epsilon$ dimensional space-time are:
\begin{align}
 \frac{m_{P_5^+}^2}{(m_x + m_y)} = &\mutwo \Big\{ 1 +
\frac{1}{\Lambdafactor\denom}\Big[\sum_j R_j^{[2,1]} (\{\cM_{XY_I}^{[2]}\})
\cR_\epsilon\mjpower \nonumber \\*
                                   &-2a^2 \deltapvtwo
\sum_{j}R_j^{[3,1]}(\{\cM_{XY_V}^{[3]}\})
                                   \cR_\epsilon\mjpower  + (V\leftrightarrow A) +
a^2(\TildeLpptwob
+ \TildeLptwob)\Big] \nonumber \\*
                                   &+\frac{\mutwo}{\Lambdafactor\ftwo^2}(4l_3^0+p_1^0+4
p_2^0)(2m_l)
+
                                     \frac{\mutwo}{\Lambdafactor\ftwo^2}(-p_1^0-4
p_2^0)(m_x
+m_y) \Big\},  \label{eq:mpisqind} \\
f_{P_5^+} = &\ftwo \Big\{ 1 + \frac{1}{\Lambdafactor\denom} \Big[
-\frac{1}{32}\sum_{Q,B}
\cR_\epsilon m^2_{Q_B}(m^2_{Q_B})^{-\frac{\epsilon}{2}} \nonumber \\*     
                                   &+ \frac{1}{4} \Big(
\cR_\epsilon\mxisq(\mxisq)^{-\frac{\epsilon}{2}} +
\cR_\epsilon\myisq(\myisq)^{-\frac{\epsilon}{2}} \nonumber \\*
                                   &+ (\muisq-\mxisq)(-\cR_\epsilon-1)(\mxisq)^{-\frac{\epsilon}{2}}
+ (\muisq -
\myisq) (-\cR_\epsilon-1)(\myisq)^{-\frac{\epsilon}{2}} \Big) \nonumber \\*
                                   &-\frac{1}{2}\Big(
R^{[2,1]}_{X_I}(\{\cM^{[2]}_{XY_I}\}) \cR_\epsilon \mxisq(\mxisq)^{-\frac{\epsilon}{2}} +
                                    R^{[2,1]}_{Y_I}(\{\cM^{[2]}_{XY_I}\})\cR_\epsilon\myisq(\myisq)^{-\frac{\epsilon}{2}}
\Big) \nonumber \\*
                                   &+\frac{a^2 \deltapvtwo}{2}\Big(
R^{[2,1]}_{X_V}(\{\cM^{[2]}_{X_V}\}) (-\cR_\epsilon-1)(\mxvsq)^{-\frac{\epsilon}{2}} +
\sum_j D^{[2,1]}_{j,
X_V}(\{\cM^{[2]}_{X_V}\}) \cR_\epsilon m_j^2(m_j^2)^{-\frac{\epsilon}{2}} \nonumber \\*
                                   &+ (X \leftrightarrow Y) + 2\sum_j
R^{[3,1]}_{j}(\{\cM^{[3]}_{XY_V}\})\cR_\epsilon m_j^2(m_j^2)^{-\frac{\epsilon}{2}} \Big)
+ (V \leftrightarrow
A)\nonumber \\*
                                   &+ a^2(\TildeLpptwob - \TildeLptwob) \Big]
+\frac{\mutwo}{2 \Lambdafactor\ftwo^2}(4l_4^0 - p_1^0)(2m_l) +
\frac{\mutwo}{2\Lambdafactor\ftwo^2}(p_1^0)(m_x+m_y) \Big\}, \label{eq:fpiind}
\end{align}
where $\Lambda$ is the scale introduced in the dimensional regularization, and
all the scale factors are written explicitly. Here, $\cR_\epsilon$ is defined to be:
\begin{equation}
\cR_\epsilon = -\frac{2}{\epsilon} - \log(4\pi) + \gamma - 1 + \cO(\epsilon),
\end{equation}
where $\gamma=-\Gamma'(1)$ is Euler's constant. In Eqs.~(\ref{eq:mpisqind}) and 
(\ref{eq:fpiind}), $\cR_\epsilon$ comes from the integral over the tadpole diagram with a 
single pole, while $(-\cR_\epsilon-1)$ comes from the integral 
over the tadpole diagram  with a double pole. The index Q runs over the 4 mesons made from one valence and one sea
quark, and B runs over the 16 tastes, which form five multiplets ($P,V,A,T,I$). ${\delta'_V}^{(2)}$ and
${\delta'_A}^{(2)}$ are LO taste-violating hairpin parameters, and $\Lpptwo$ and
$\Lptwo$ are NLO
taste-violating parameters. The latter are simply the linear combinations of
LECs coming from $\cO(a^2 p^2)$ and $\cO(a^2 m_q)$ taste-violating terms, for
example, the operators given in Eq.~(\ref{eq:NLOtvterms}). There are
no contributions from $\cO(a^4)$ terms to pseudo-Goldstone masses and decay constants, either because
of the exact non-singlet chiral symmetry (for the masses) or because the operators do
not contain derivatives (for the decay constant).

The residue functions R and D are defined as in the SU(3) case~\cite{AB_M}:
\begin{align}
        R^{[n,k]}_j (\{M\};\{\mu\}) & \equiv \frac{\Pi^k_{a=1}(\mu_a^2 -
m_j^2)}{{\Pi'}^{n}_{l=1} (m_l^2 - m_j^2)},  \label{eq:R} \\*
        D^{[n,k]}_{j,i} (\{M\};\{\mu\})  & \equiv
-\frac{d}{dm_i^2}R^{[n,k]}_j(\{M\};\{\mu\}), \label{eq:D}
\end{align}
where the prime on the product means that $l=j$ is omitted. The denominator mass-set 
arguments in Eqs.~(\ref{eq:R}) and (\ref{eq:D}) are defined by:
\begin{align}
\{\cM_{X_V}^{[2]}\} & \equiv \{ m_{X_V}, m_{\eta'_V} \},
&\{\cM_{Y_V}^{[2]}\}  &\equiv \{ m_{Y_V}, m_{\eta'_V} \}, \nonumber \\*
\{\cM_{XY_I}^{[2]}\} & \equiv \{ m_{X_I}, m_{Y_I} \},
&\{\cM_{XY_V}^{[3]}\}  &\equiv \{ m_{X_V}, m_{Y_V}, m_{\eta'_V} \}.
\end{align}
The numerator mass-set arguments for taste $\Xi$ are always $\{\mu_\Xi\} \equiv
\{m_{U_\Xi}\}$.
We show the masses explicitly here:
\begin{align}       
        m^2_{\pi_B}  &= m^2_{U_B} = m^2_{D_B} = 2\mutwo m_l +
a^2\Delta_B^{(2)}, \label{eq:mpiB} \\
        m^2_{X_B} &= 2 \mutwo m_{x} + a^2\Delta_B^{(2)}, \label{eq:mxB} \\
        m^2_{Y_B} &= 2 \mutwo m_{y} + a^2\Delta_B^{(2)}, \label{eq:myB} \\
        m^2_{\eta'_V} &= \muvsq + \frac{a^2 \deltapvtwo}{2}, \label{eq:metapV}
\\
        m^2_{\eta'_A} &= \muasq + \frac{a^2 \deltapatwo}{2}, \label{eq:metapA}
\\
        m^2_{\eta'_I} & \sim \frac{2}{3}m_0^2,  \label{eq:metapI}
\end{align}
where $\Delta_B^{(2)}$ are the taste splittings in SU(2) \rschpt. The final
relation
holds for $m_0^2 \gg m_{\pi_I}^2$. Here, $\eta'_V$ and $\eta'_A$ are,
respectively,
the taste-vector and taste-axial vector, flavor and replica neutral mesons whose
masses are shifted by the taste-violating hairpin contributions.
Since $\eta'_I$ has a mass proportional to $m_0^2$, it decouples
in the limit when $m_0^2$ is taken to infinity. 

Using the identities of residue functions listed in the second paper of
Ref.~\cite{AB_M}:
\begin{eqnarray}\label{eq:identities}
\sum_{j=1}^n  R_{j}^{[n,k]} &=& \begin{cases} \phantom{-}1\ , &n=k+1 ;\\
                     \phantom{-}0\ , &n\ge k+2 .
\end{cases} \nonumber \\*
\sum_{j=1}^n  R_{j}^{[n,k]}m^2_j  &= & \begin{cases} \sum_{j=1}^n m^2_j -
                \sum_{a=1}^k \mu^2_a  \ , &n=k+1 ; \\
                     -1\ , &n=k+2 ; \\
                     \phantom{-}0\ , &n\ge k+3 .
\end{cases} \nonumber \\*
\sum_{j=1}^n  D_{j,\ell}^{[n,k]} &=& \begin{cases} \phantom{-}1\ , &n=k ; \\
                     \phantom{-}0\ , &n\ge k+1 .
\end{cases}\nonumber \\*
\sum_{j=1}^n  \left(D_{j,\ell}^{[n,k]}m^2_j\right)
- R_{\ell}^{[n,k]}  &= & \begin{cases} m^2_\ell + \sum_{j=1}^n m^2_j -
                \sum_{a=1}^k \mu^2_a  \ , &n=k ; \\
                     -1\ , &n=k+1 ; \\
                      \phantom{-}0\ , &n\ge k+2 .\end{cases}
\end{eqnarray}
and ignoring terms vanishing at order $\epsilon$ or higher as $\epsilon\to 0$ , 
one can simplify Eqs.~(\ref{eq:mpisqind}) and (\ref{eq:fpiind}) to:

\begin{align}
 \frac{m_{P_5^+}^2}{(m_x + m_y)} = &\mutwo \Big\{ 1 +
\frac{1}{\denom}\Big[ (\mutwo(2m_x + 2m_y - 2m_l) + a^2\Delta_I^{(2)}+2a^2\deltapvtwo +
2a^2\deltapatwo)\cR_\epsilon \nonumber \\*
& + \sum_j R_j^{[2,1]}(\{\cM_{XY_I}^{[2]}\}) l(m_j^2) -2a^2 \deltapvtwo
\sum_{j}R_j^{[3,1]}(\{\cM_{XY_V}^{[3]}\})
                                   l(m_j^2) + (V\leftrightarrow A)\nonumber \\*
		&+ \Lambdafactor a^2(\TildeLpptwob + \TildeLptwob)\Big] \nonumber \\*
               	&+\frac{\mutwo}{\Lambdafactor\ftwo^2}(4l_3^0+p_1^0+4p_2^0)(2 m_l)
               	+\frac{\mutwo}{\Lambdafactor\ftwo^2}(-p_1^0-4p_2^0)(m_x
             	+m_y) \Big\},  \label{eq:mpisqindsim} \\
f_{P_5^+} = &\ftwo \Big\{ 1 + \frac{1}{\denom} \Big[
		-(\mutwo(m_x+m_y+2m_l)+2a^2\Delta_{av}^{(2)}+2a^2\deltapvtwo 
		+2a^2\deltapatwo)\cR_\epsilon \nonumber \\*
		& -\frac{1}{32}\sum_{Q,B}l(m_{Q_B}^2) + \frac{1}{4} \Big(
		  l(\mxisq) + l(\myisq) + (\muisq-\mxisq)\tilde l(\mxisq) \nonumber \\*
		&+ (\muisq - \myisq) \tilde l(\myisq) \Big) -\frac{1}{2}\sum_{j}R^{[2,1]}_{m_j}(\{\cM^{[2]}_{XY_I}\})l(m_j^2) \nonumber \\*
          	&+\frac{a^2 \deltapvtwo}{2}\Big(R^{[2,1]}_{X_V}(\{\cM^{[2]}_{X_V}\}) \tilde l(\mxvsq) 
		 + \sum_j D^{[2,1]}_{j, X_V}(\{\cM^{[2]}_{X_V}\}) l(m^2_j) \nonumber \\*
               	&+ (X \leftrightarrow Y)  + 2\sum_j R^{[3,1]}_{j}(\{\cM^{[3]}_{XY_V}\})l(m_j^2)\Big) 
		 + (V \leftrightarrow A)\nonumber \\*
              	&+ \Lambdafactor a^2(\TildeLpptwob - \TildeLptwob) \Big]
		 +\frac{\mutwo}{2 \Lambdafactor\ftwo^2}(4l_4^0 - p_1^0)(2 m_l) 
		 + \frac{\mutwo}{2\Lambdafactor\ftwo^2}(p_1^0)(m_x+m_y) \Big\}, \label{eq:fpiindsim}
\end{align}
where
\begin{equation}
\Delta_{av}^{(2)} \equiv \frac{1}{16}(\Delta_5^{(2)} + 4 \Delta_V^{(2)} + 6
\Delta_T^{(2)} + 4 \Delta_A^{(2)} + \Delta_I^{(2)})
\end{equation}
is the average taste splitting in the two-flavor case.
The chiral logarithm functions $l$ and $\tilde l$ in Eqs.~(\ref{eq:mpisqindsim})
and
(\ref{eq:fpiindsim}) are given by \cite{AB_M}:
\begin{align}
l(m^2) &\equiv m^2 \ln \frac{m^2}{\Lambda^2} \qquad [\text{infinite volume}],
\label{eq:l} \\
\tilde l(m^2) &\equiv -\left(\ln\frac{m^2}{\Lambda^2} + 1\right) \qquad
[\text{infinite volume}]. \label{eq:tildel}
\end{align}
Finite volume corrections at NLO may be incorporated by adjusting $l(m^2)$ and 
$\tilde l(m^2)$ as in Ref.~\cite{AB_M, CB_PRD}.

Recall that in continuum SU(2) \chpt, because the NLO Lagrangian contains all the
possible analytic terms consistent with the symmetries, the divergences 
generated from one-loop graphs built from LO vertices can be absorbed by an appropriate renormalization of the
bare NLO LECs $l_i^0$ and contact term coefficients $h_i^0$~\cite{GL_SU2}:
\begin{align}
l_i^0 &= (\Lambda)^{d-4} (l_i + \gamma_i \frac{\cR_\epsilon}{32\pi^2}), \ i=1,\cdots, 7,
\label{eq:renorm_li}\\
h_i^0 &= (\Lambda)^{d-4} (h_i + \delta_i\frac{\cR_\epsilon}{32\pi^2}), \ h=1, 2, 3,
\label{eq:renorm_hi} 
\end{align}
where $\cR_\epsilon$ has the same definition as above, and $l_i$ and $h_i$ are renormalized
coefficients (which often appear as $l_i^r$ and $h_i^r$ in literature). For SU(2) \chpt, the values of $\gamma_i$
and $\delta_i$ are listed in Ref.~\cite{GL_SU2}. For the general case in SU(N) \chpt, 
similar results can be found in Ref.~\cite{BIJNENS_RENORM}. In
Eqs.~(\ref{eq:renorm_li}) and (\ref{eq:renorm_hi}), as one changes the scale
$\Lambda$, $l_i$ and $h_i$ should also change in such a way that the bare quantities
$l_i^0$ and $h_i^0$ are scale independent. Specifically, under a change in the chiral 
scale $\Lambda$ to $\Lambda'$, the SU(2) LECs change by:

\begin{equation}
l_i(\Lambda') = l_i(\Lambda) - \frac{\gamma_i}{32\pi^2}
\log\frac{{\Lambda'}^2}{\Lambda^2},
\end{equation}

This renormalization procedure can be applied in SU(2) \rschpt\ in the same way. 
The only difference is that, at each order of chiral expansion, there are additional taste-violating
terms. The presence of these terms in effective field theory reflects the fact
that the continuum SU(4) taste symmetry is broken by finite lattice spacing effects. 
In the two-flavor case, the full chiral symmetry $SU_L(8)\times SU_R(8)$ is
broken both by taste-violating terms and by the usual mass terms. Effectively, the taste-violating terms are
acting just like the mass terms, and they can be treated in the same way once
the power counting scheme is specified. In practice, we use the power
counting rule $p^2 \sim m_q \sim a^2$ in SU(2) \rschpt ~\cite{AB_M, MILC_REVIEW}.
As a result, the LO contribution for a physical quantity is at $\cO(p^2),
\cO(m_q)$ and $\cO(a^2)$, coming from the terms in Eq.~(\ref{eq:LagLO}). At NLO, 
the one-loop graphs built from LO vertices will generate divergences at
$\cO(p^4), \cO(p^2m_q), \cO(m_q^2), \cO(a^2p^2), \cO(a^2 m_q)$ and $\cO(a^4)$.
By construction, Eq.~(\ref{eq:LagNLO}) is the most general Lagrangian in the
same order which satisfies all the symmetries of staggered quarks. Indeed, all
possible terms in this Lagrangian are found by treating mass terms and
taste-violating terms in the same footing, using a spurion
analysis~\cite{MILC_REVIEW}. Since the staggered symmetries (a subset of
$SU_L(8)\times SU_R(8)$ in the two-flavor case) are not violated by dimensional
regularization, it is possible to absorb all the one-loop divergences by renormalization of the 
NLO LECs in $\cL_{cont}^{(4)}$ and NLO taste-violating parameters in $\cL_{\operatorname{t-v}}^{(4)}$. 
This is indeed the case in current calculations of the pseudo-Goldstone pion mass and decay constant. 
However, since I am only concentrating on these two physical quantities, 
I can only derive the renormalization conditions for certain linear combinations 
of LECs and taste-violating parameters.
Since valence quark masses $m_x, m_y$, sea quark mass $m_l$ and lattice spacing $a^2$
each can vary independently, one can collect the coefficients for each term
separately and obtain the following renormalizations:
\begin{align}
l_3^0 &= \Lambda^{d-4} (l_3 -
\frac{1}{64\pi^2}\cR_\epsilon),\label{eq:renorm_l3}\\
l_4^0 &= \Lambda^{d-4} (l_4 + \Cc \cR_\epsilon),\label{eq:renorm_l4}\\
p_1^0 &= \Lambda^{d-4} (p_1 + \frac{1}{8\pi^2}
\cR_\epsilon),\label{eq:renorm_p1}\\
p_2^0 &= \Lambda^{d-4} p_2, \label{eq:renorm_p2}\\
(\TildeLpptwob + \TildeLptwob) &= \Lambda^{d-4} (\TildeLpptwo + \TildeLptwo -
(\Delta_I + 2\deltapv + 2\deltapa) \cR_\epsilon),\label{eq:renorm_tp}\\
(\TildeLpptwob - \TildeLptwob) &= \Lambda^{d-4} (\TildeLpptwo - \TildeLptwo +
2(\Delta_{av} +
\deltapv + \deltapa) \cR_\epsilon), \label{eq:renorm_tm}
\end{align}
Again, the renormalized coupling constants in SU(2) \schpt\ are scale dependent.
They should change with the scale $\Lambda$ in such a way that the bare
coefficients are scale independent. It is easily seen from
Eqs.~(\ref{eq:renorm_l3})-(\ref{eq:renorm_tm}) that, under a change in the
chiral scale $\Lambda$ to $\Lambda'$, the LECs change by:
\begin{align}
l_3(\Lambda') &= l_3(\Lambda)
+ \frac{1}{64\pi^2} \log\frac{{\Lambda'}^2}{\Lambda^2},  \\
l_4(\Lambda') &= l_4(\Lambda) - \Cc \log\frac{{\Lambda'}^2}{\Lambda^2}, \\
p_1(\Lambda') &= p_1(\Lambda) - \frac{1}{8\pi^2} \log\frac{{\Lambda'}^2}{\Lambda^2}, \\
p_2(\Lambda') &= p_2(\Lambda), \\
(\TildeLpptwo + \TildeLptwo)(\Lambda') &= (\TildeLpptwo + \TildeLptwo)(\Lambda)
+ (\Delta_I + 2\deltapv + 2\deltapa)\log\frac{{\Lambda'}^2}{\Lambda^2},  \\
(\TildeLpptwo - \TildeLptwo)(\Lambda') &= (\TildeLpptwo - \TildeLptwo)(\Lambda)
- 2(\Delta_{av} + \deltapv + \deltapa)\log\frac{{\Lambda'}^2}{\Lambda^2}.
\end{align}

After the renormalizations in Eq.~(\ref{eq:renorm_l3}) through
Eq.~(\ref{eq:renorm_tm}), the pion mass and decay constant can be written in
terms of renormalized LECs and taste-violating parameters:
\begin{align}
 \frac{m_{P_5^+}^2}{(m_x + m_y)} = &\mutwo \Big\{ 1 +
\frac{1}{\denom}\Big[\sum_j R_j^{[2,1]} (\{\cM_{XY_I}^{[2]}\}) l(m_j^2)\nonumber \\*
		&- 2a^2 \deltapvtwo\sum_{j}R_j^{[3,1]}(\{\cM_{XY_V}^{[3]}\})
                         l(m_j^2) + (V\leftrightarrow A) + a^2(\TildeLpptwo +
\TildeLptwo)\Big] \nonumber \\*
               	& + \frac{\mutwo}{\ftwo^2}(4l_3+p_1+4p_2)(m_u+m_d) +
                  \frac{\mutwo}{\ftwo^2}(-p_1-4 p_2)(m_x + m_y) \Big\},  \label{eq:mpisq} \\
f_{P_5^+} = &\ftwo \Big\{ 1 + \frac{1}{\denom} \Big[ -\frac{1}{32}\sum_{Q,B}l(m_{Q_B}^2) \nonumber \\*     
                       &+ \frac{1}{4} \Big( l(\mxisq) + l(\myisq) +
                    (\muisq-\mxisq)\tilde l(\mxisq) + (\muisq -
\myisq) \tilde l(\myisq) \Big) \nonumber \\*
                       &- \frac{1}{2}\Big(R^{[2,1]}_{X_I}(\{\cM^{[2]}_{XY_I}\}) l(\mxisq) 
                        + R^{[2,1]}_{Y_I}(\{\cM^{[2]}_{XY_I}\})l(\myisq)\Big) \nonumber \\*
                         &+\frac{a^2 \deltapvtwo}{2}\Big(
R^{[2,1]}_{X_V}(\{\cM^{[2]}_{X_V}\}) \tilde l(\mxvsq) + \sum_j D^{[2,1]}_{j,
X_V}(\{\cM^{[2]}_{X_V}\}) l(m^2_j) \nonumber \\*
                                   &+ (X \leftrightarrow Y) + 2\sum_j
R^{[3,1]}_{j}(\{\cM^{[3]}_{XY_V}\})l(m_j^2)\Big) + (V \leftrightarrow
A)\nonumber \\*
                       &+ a^2(\TildeLpptwo - \TildeLptwo) \Big]
+\frac{\mutwo}{2 \ftwo^2}(4l_4 - p_1)(m_u+m_d) + \frac{\mutwo}{2
\ftwo^2}(p_1)(m_x+m_y) \Big\}. \label{eq:fpi}
\end{align}

Now that we have the results for the pion mass and decay constant to NLO in
SU(2) \pqrschpt, we can study the relations of the LECs and taste-violating parameters
between the two-flavor and the three-flavor cases. This can be done by comparing
formulae for physical quantities in SU(2) theory and the corresponding formulae in SU(3)
theory, in the case where the light quark masses and taste splittings are much
smaller than the strange quark mass, \ie
\begin{equation}
\frac{m_x}{m_s}, \frac{m_y}{m_s}, \frac{m_l}{m_s},
\frac{a^2\Delta_B}{\mu m_s}, \frac{a^2\delta'_{V(A)}}{\mu m_s} \sim
\epsilon \ll 1. \label{eq:su2limit}
\end{equation}

For small $\epsilon$, we expect the SU(2) theory to be generated from the 
SU(3) one as in Ref.~\cite{GL_SU3}. Since, at NLO in SU(3) \chpt,
there are terms which go like $\frac{\mu m_s}{(4\pi f)^2}$ times logarithms, we
will in general need to expand to $\cO(\epsilon)$ to pick up all terms that
appear at NLO in SU(2) \chpt, such as $\frac{\mu m_l}{(4\pi f)^2}$ or $
\frac{a^2\Delta_B}{(4\pi f)^2}$. Of course, all dependence on $m_x, m_y, m_l$
and $a^2$ must be explicit, because the LECs do not depend on the light quark 
masses and have no power-law dependence on lattice spacings.

I will first focus on the taste-splittings $\Delta^{(2)}_B$ and the
taste-violating hairpin
parameters $\delta'^{(2)}_{V(A)}$. In Eqs.~(\ref{eq:mpisq}) and (\ref{eq:fpi}),
$\Delta^{(2)}_B$ and $\delta'^{(2)}_{V(A)}$ only appear in the NLO part, and the
same statement is true for $\Delta_B$ and $\delta'_{V(A)}$ in the corresponding
SU(3) formulae, so it suffices to use the relations between $\Delta^{(2)}_B$ and
$\Delta_B$,
and $\delta'^{(2)}_{V(A)}$ and $\delta'_{V(A)}$, at LO in \rschpt.

At LO in SU(3) \rschpt, we have the mass of a flavor-nonsinglet meson:
\begin{equation}
m^2_{U_B} = 2\mu m_l + a^2\Delta_B. \label{eq:muB_SU3}
\end{equation}
By comparing with Eq.~(\ref{eq:mpiB}), we conclude that at LO, for each taste
index $B$, we have
\begin{equation}
a^2\Delta_B^{(2)} = a^2\Delta_B. \label{eq:Delta}
\end{equation}

On the other hand, the mass of $\eta_V$, the lighter of the two flavor-neutral,
taste-vector mesons that mix in the SU(3) \rschpt\, is:
\begin{align}
m^2_{\eta_V} &= \frac{1}{2}\left(m^2_{U_V} + m^2_{S_V} + \frac{3}{4}a^2\deltapv
-Z\right), \\
        Z    &= \sqrt{(m^2_{S_V}- m^2_{U_V})^2 - \frac{a^2\deltapv}{2}(m^2_{S_V}
- m^2_{U_V}) + \frac{9(a^2\deltapv)^2}{16}}.
\end{align}
In the limit $m_l = m_u \ll m_s$ and $a^2\delta'_{V(A)} \ll \mu m_s$, it should
become
the mass of what we call $\eta'_V$ here, as given in Eq.~(\ref{eq:metapV}).
Indeed, we have:
\begin{align}
m^2_{\eta_V} &  \underset{m_l \ll m_s}{\longrightarrow} m^2_{U_V} +
\frac{1}{2}a^2\deltapv + \cO(\frac{(a^2\deltapv)^2}{\mu m_s}). \label{eq:metaV}
\end{align}
Comparing Eq.~(\ref{eq:metapV}) and Eq.~(\ref{eq:metaV}), we find that, at LO in
\rschpt,
\begin{align}
a^2\deltapvtwo  = a^2 \deltapv,
\label{eq:deltapvtwo}
\end{align}
where corrections of $\cO(\frac{(a^2\deltapv)^2}{\mu m_s})$ generate NLO effects in
SU(2) \rschpt, since they are of $\cO(a^4)$.
A similar relation holds for $\deltapatwo$ and $\deltapa$ at LO:
\begin{equation}
a^2\deltapatwo = a^2\deltapa. \label{eq:deltapatwo}
\end{equation}

If we expand the NLO SU(3) formulae for $m_\pi^2$ and $\fpi$ in Ref.~\cite{AB_M}
in powers of $\epsilon$, we find that the three-flavor formulae reproduce the
form of the two-flavor formulae, as expected. Both are
expansions in orders of $m_x, m_y, m_l, a^2\Delta_B$ and $a^2\delta'_{V(A)}$.
Since the light valence quark masses, sea quark masses and lattice spacings can vary independently, 
we can match the coefficient of each term. 
By comparing formulae in SU(2) \schpt\ and SU(3) \schpt,
and utilizing Eqs.~(\ref{eq:Delta}), (\ref{eq:deltapvtwo}) and
(\ref{eq:deltapatwo}), one obtains the relations between SU(2) LECs and SU(3)
LECs up to NLO. I find:
\begin{align}
\ftwo &= f (1 - \frac{1}{16\pi^2 f^2} \mu m_s \log\frac{\mu m_s}{\Lambda^2} +
\frac{16L_4}{f^2}\mu m_s), \label{eq:ftwo} \\
\mutwo &= \mu (1 - \frac{1}{48\pi^2 f^2} \frac{4\mu m_s}{3}\log\frac{\frac{4\mu
m_s}{3}}{\Lambda^2} + \frac{32(2L_6-L_4)}{f^2}\mu m_s), \label{eq:mutwo} \\
p_1 &= 16 L_5 - \Cc (1 + \logmk), \label{eq:p1}  \\
p_2 &= -8 L_8 + \Cc \frac{1}{6}(\logmeta) + \Cc \frac{1}{4}(1 + \logmk),
\label{eq:p2} \\
l_3 &= 8(2L_6 - L_4) + 4(2L_8 - L_5) - \Cc \frac{1}{36}(1 + \logmeta),
\label{eq:l3}  \\
l_4 &= 8L_4 + 4L_5 - \Cc \frac{1}{4} (1 + \logmk), \label{eq:l4} \\
\TildeLpptwo &= \tilde L'' - \frac{1}{6}\Delta_I(1+\logmeta)
-\frac{1}{2}\Delta_{av}(1+\logmk), \label{eq:Lpptwo} \\
\TildeLptwo &= \tilde L' - \frac{1}{6}\Delta_I(1+\logmeta)
+\frac{1}{2}\Delta_{av}(1+\logmk), \label{eq:Lptwo}
\end{align}
where $L_4, L_5, L_6$ and $L_8$ are renormalized SU(3) LECs, $\tilde L''$ and $\tilde L'$ 
are the NLO taste-violating parameters in SU(3) \rschpt. Here I use the tilde to
distinguish them from $L''$ and $L'$ after redefinitions in Ref.~\cite{MILC_04}.
Namely, in SU(3) \rschpt, $L''$ and $L'$ are related to $\tilde L''$ and $\tilde L'$ through
\begin{align}
\Cc(L'' - L') &= \Cc (\tilde L'' - \tilde L') - (8 L_5 + 24 L_4)
\Delta_{av}^{(2)},  \label{EQ:LREDEFSU3_1} \\
\Cc(L'' + L') &= \Cc (\tilde L'' + \tilde L') - (32L_8- 16L_5 +  96L_6 -
48L_4)\Delta_I^{(2)}. \label{EQ:LREDEFSU3_2}
\end{align}
Eqs. (\ref{eq:l3}) and (\ref{eq:l4}) are the same as the equations in the full
QCD continuum case~\cite{GL_SU3}. Eqs. (\ref{eq:p1}) and (\ref{eq:p2}) relate
the unphysical LECs in the
partially-quenched two-flavor theory to the physical LECs in the three-flavor
theory. Eqs. (\ref{eq:Lpptwo}) and (\ref{eq:Lptwo}) give us relations between
taste-violating parameters in the two-flavor and three-flavor theories. If we
require the SU(2) \schpt\ to describe the same physics in the two-flavor sector
of the underlying SU(3) \schpt, all the parameters in the SU(2) theory should
vary with the strange quark mass $m_s$ according to
Eqs.~(\ref{eq:p1})-(\ref{eq:Lptwo}).

The renormalizations of $\TildeLpptwo$ and $\TildeLptwo$ are complicated and
involve the taste-splitting terms $\Delta_I$ and $\Delta_{av}$. It is more
convenient to redefine $\TildeLpptwo$
and $\TildeLptwo$ by associating particular $O(a^2)$ terms with the
$l_i$~\cite{MILC_04}.
The following replacements:
\begin{align}
\frac{\mutwo}{2\ftwo^2}(p_1)(m_u + m_d) &\to \frac{p_1}{2\ftwo^2} (\mutwo(m_u + m_d) +
a^2\Delta_{av}^{(2)}), \nonumber \\
\frac{\mutwo}{2\ftwo^2}(4l_4 - p_1)(m_x + m_y) &\to \frac{4l_4 - p_1}{2\ftwo^2}
(\mutwo(m_x + m_y) + a^2\Delta_{av}^{(2)}), \nonumber \\
\frac{\mutwo}{\ftwo^2}(4l_3 + p_1 + 4p_2)(m_u + m_d) &\to
\frac{4l_3+p_1+4p_2}{\ftwo^2} (\mutwo(m_u + m_d) + a^2\Delta_I^{(2)}), \nonumber \\
\frac{\mutwo}{\ftwo^2}(-(p_1 + 4p_2))(m_x + m_y) &\to
\frac{-(p_1+4p_2)}{\ftwo^2} (\mutwo(m_x + m_y) + a^2\Delta_I^{(2)}) \label{eq:lredef}
\end{align}
absorb splittings into the mass-dependent counterterms to make them
correspond to the meson masses (or average values thereof) that appear in the
loops. Eq.~(\ref{eq:lredef}) is equivalent to defining new parameters $\Lpptwo$
and
$\Lptwo$:
\begin{align}
\Cc (\Lpptwo - \Lptwo) &= \Cc (\TildeLpptwo - \TildeLptwo) + 2l_4
\Delta_{av}^{(2)}, \\
\Cc (\Lpptwo + \Lptwo) &= \Cc (\TildeLpptwo + \TildeLptwo) - 4 l_3
\Delta_{I}^{(2)}.
\end{align}
After these redefinitions, $\Lpptwo$ will become independent of chiral
scale, and $\Lptwo$ is renormalized according to:
\begin{align}
\Lptwo(\Lambda') = \Lptwo(\Lambda) + 2(\deltapvtwo + \deltapatwo)\log
\frac{{\Lambda'}^2}{\Lambda^2}.
\end{align}
The renormalizations of other LECs remain unchanged.

After these redefinitions, the new $\Lpptwo$ and $\Lptwo$ are related to the
corresponding SU(3) quantities $L''$ and $L'$ by:
\begin{align}
\Lpptwo - \Lptwo &= (L'' - L') - \Delta_{av}(1+\logmk) + 16\pi^2\Delta_{av}
(8L_5 + 24 L_4 -
2l_4) \nonumber \\
         &= (L'' - L') + \Delta_{av}\left[128\pi^2L_4 -
\frac{1}{2}(1+\logmk) \right] \label{eq:lppmlp} \\
\Lpptwo + \Lptwo &= (L'' + L') -\frac{1}{3}\Delta_I(1 + \logmeta) +
16 \pi^2 \Delta_I (32L_8 - 16L_5 + 96L_6 - 48L_4 - 4l_3) \nonumber \\
        &= (L'' + L') + \Delta_I \left[16\pi^2(32L_6 - 16L_4)
- \frac{2}{9}(1+\logmeta) \right] \label{eq:lppplp}
\end{align}
Using the standard scale renormalization of the $L_i$~\cite{GL_SU3},
\begin{align}
L_i(\Lambda') = L_i(\Lambda) +
\frac{C_i}{256\pi^2}\log\frac{{\Lambda'}^2}{\Lambda^2}
\end{align}
with
\begin{eqnarray}\label{eq:Ci}
C_4=-1\;; &\qquad& C_5= -3\;;\\
2C_6-C_4=-2/9\;; &\qquad& 2C_8-C_5= 4/3,
\end{eqnarray}
it is easy to check that the factors in square parenthesis in Eqs.
(\ref{eq:lppmlp}) and (\ref{eq:lppplp}) are scale independent. This is a
consistency check, since $\Lpptwo$ and $\Lptwo$ transform in the same way
as $L''$ and $L'$, respectively, under scale change.

\section{REMARKS AND CONCLUSION}

I calculated the pseudo-Goldstone pion mass and decay constant to NLO in
two-flavor \pqrschpt\, using the replica method. I also checked that SU(2)
\rschpt\ emerges from SU(3) \rschpt\ in the limit $\frac{m_x}{m_s},
\frac{m_y}{m_s},\frac{m_l}{m_s},\frac{a^2\Delta_B}{\mu
m_s},\frac{a^2\delta'_{V(A)}}{\mu m_s} \ll 1$, as assumed in
Ref.~\cite{CB_SCHPT}. Finally, I derived the relations for
the LECs and taste-violating parameters between the two-flavor and three-flavor
cases. Some of the formulae here (Eqs.~(\ref{eq:mpisq}) and (\ref{eq:fpi})) are
used for
the SU(2) chiral fits to MILC data~\cite{DU_LAT09}.

At the present stage, we have MILC data for the light pseudoscalar mass and
decay constant at five lattice spacings from 0.15\,fm to 0.045\,fm, generated
with 2+1
flavors of asqtad improved staggered quarks.
For each lattice spacing, we have many different sea quark masses as well as
many different
combinations of valence quark masses. For most ensembles, the strange quark mass
is near its physical value, and the light sea quark masses are much smaller.
If light valence quark masses and taste splittings are also taken significantly
smaller
than the strange quark mass, we expect that SU(2) \rschpt\ would apply.
Preliminary results
indicate that it is indeed the case. Since the strange quark mass is close to
the physical
value in the ensembles used for the fits, the SU(2) LECs only suffer small
changes
due to variations in the strange quark mass. We can fit to lattice data using
Eqs.~(\ref{eq:mpisq})
and (\ref{eq:fpi}) to get values of SU(2) LECs, the pion decay constant $f_\pi$,
and the physical light quark mass $\hat m$, as well as the chiral condensate in
the
two-flavor chiral limit. Furthermore, we can do a systematic NNLO SU(2) chiral
fit if continuum NNLO chiral logarithms~\cite{BIJ_SU2} and possible analytic
terms are
included, and if taste-violations are relatively small. The results appear to be
consistent
with the results of the SU(3) analysis~\cite{DU_LAT09}.

However, to make the formulae complete and results more accurate, it may be
important to incorporate the effects of the variations in the strange quark
mass by doing appropriate adjustments on certain parameters in the two-flavor
theory.
In practice, for each strange quark mass, the four LECs $l_3,l_4,p_1$ and $p_2$
may be adjusted according to Eqs.~(\ref{eq:p1})-(\ref{eq:l4}), and the two
taste-violating
parameters, $\Lpptwo$ and $\Lptwo$, may be adjusted according to
Eq.~(\ref{eq:lppmlp}) and Eq.~(\ref{eq:lppplp}). One then performs chiral fits to
all the lattice data simultaneously. At the final step, physical values of LECs
can be obtained by extrapolating to the physical strange quark mass.

An extension of the present work to the case of quantities involving the strange
quark such as $f_K$ or $m^2_K$ using the method of heavy kaon
\chpt~\cite{HEAVYs_RBC, ROESSL} may be very useful. Work on that is in progress.

\section*{ACKNOWLEDGEMENT}
I thank C.\ Bernard for proposing this subject and for valuable advice along the
way.

\end{document}